\documentclass{article}

\usepackage{PRIMEarxiv}

\usepackage[utf8]{inputenc} 
\usepackage[T1]{fontenc}    
\usepackage{hyperref}       
\usepackage{url}            
\usepackage{booktabs}       
\usepackage{amsfonts}       
\usepackage{nicefrac}       
\usepackage{microtype}      
\usepackage{lipsum}
\usepackage{fancyhdr}       
\usepackage{graphicx}       
\graphicspath{{media/}}     
\usepackage{amsmath,amssymb}
\usepackage{subcaption}
\usepackage{caption}
\usepackage{booktabs} 
\usepackage{siunitx} 
\usepackage{amsmath} 
\usepackage{cite}
\usepackage{graphicx}
\usepackage{caption}
\usepackage{multirow}
\usepackage{amsmath}  
\usepackage{algorithm}  
\usepackage{algpseudocode}  
\begin{document}

\title{Markowitz Meets Bellman: Knowledge-distilled Reinforcement Learning for Portfolio Management
}

\author{
  Gang Hu \\ 
  School of Computer Science \\
  Georgia Institute of Technology \\
  Atlanta\\
  \texttt{\ ghu70@gatech.edu} \\
  \and 
  Ming Gu \\
  Department of Electrical and Electronic Engineering \\
  University of Manchester \\
  Manchester, UK \\
  \texttt{ming.gu@manchester.ac.uk} \\
}
\maketitle

\begin{abstract}
Investment portfolios, central to finance, balance potential returns and risks. This paper introduces a hybrid approach combining Markowitz's portfolio theory with reinforcement learning, utilizing knowledge distillation for training agents. In particular, our proposed method, called KDD (Knowledge Distillation DDPG), consist of two training stages: supervised and reinforcement learning stages. The trained agents optimize portfolio assembly. A comparative analysis against standard financial models and AI frameworks, using metrics like returns, the Sharpe ratio, and nine evaluation indices, reveals our model's superiority. It notably achieves the highest yield and Sharpe ratio of 2.03, ensuring top profitability with the lowest risk in comparable return scenarios.
\end{abstract}


The integration of advanced AI techniques, particularly Reinforcement Learning (RL), has notably advanced the field, with significant contributions from DeepMind. Pioneering applications of RL, such as in Atari 2600 video games \cite{mnih2015human}, and the development of the AlphaGo series \cite{silver2016mastering, silver2017mastering}, have demonstrated superhuman performance, showcasing the potential of RL in complex tasks. This approach has been extended to various sectors, including robotics \cite{lillicrap2016continuous}, autonomous vehicles \cite{wang2019deep}, and finance, where AI and ML integration have enhanced profitability, as evidenced by hedge funds' performance \cite{grobys2022hedge} and the development of cost-effective financial models \cite{kolm2019deep}. These advancements underscore the transformative impact and versatility of RL in modern AI applications.

In the specialized domain of finance, RL's capabilities in handling dynamic and complex decision environments are paramount. The evolution of RL transcends traditional boundaries, notably infiltrating the domain of Online Portfolio Selection (OLPS) and traditional portfolio management. In OLPS, RL contributes to dynamic asset allocation, constantly adjusting to market changes to outperform benchmarks \cite{velay2023benchmarking}. In broader portfolio management, RL intersects with established strategies like "Follow-the-Winner," "Follow-the-Loser," "Pattern-Matching," and "Meta-Learning" \cite{li2014online,li2012pamr,cover1996universal,gyorfi2006nonparametric,vovk1998universal,das2011meta}. These strategies are now being refined and enhanced through RL's ability to adapt to and learn from complex, dynamic environments, reinforcing its role as a pivotal force in revolutionizing sectors traditionally governed by heuristic methods.

Despite RL's theoretical advancements in finance, its practical application, especially in financial trading's price prediction models, encounters specific challenges. Traditional deep learning methods in financial trading primarily predict asset price movements using historical data \cite{heaton2016deep,niaki2013forecasting,freitas2009prediction,hu2023advancing}. However, the utility of these predictions hinges on their accuracy and the subsequent transformation into viable strategies, often requiring manual intervention that hampers adaptability and scalability. Furthermore, these models frequently overlook critical factors like transaction costs, thus limiting their practicality in real-world trading scenarios.

RL-based models emerge as a more comprehensive approach, not just predicting prices but also offering a robust framework for decision-making in the complex landscape of portfolio management. Model-free, machine-learning-driven solutions in algorithmic trading align with the RL framework, eschewing reliance on precise future price predictions \cite{moody2001learning,dempster2006automated,cumming2015investigation,deng2017deep}. While RL-based models excel in generating trading signals for individual assets, they are predominantly tailored to single-asset transactions. This specialization reveals certain limitations in the multifaceted field of portfolio management, which requires a holistic strategy for managing multiple assets. The inherent challenge in this domain stems from the necessity to integrate diverse assets into a cohesive portfolio strategy."

The demand for a comprehensive portfolio management strategy highlights the potential of RL-based models to address challenges beyond single-asset trading. RL's success in discrete trading decisions suggests its capability to manage the continuous action spaces in portfolio management. Despite new challenges, RL's track record in complex scenarios suggests promising applications in advanced portfolio strategies. RL excels in discrete action spaces such as video gaming \cite{mnih2015human} and board games \cite{silver2016mastering}, but faces challenges in the continuous action spaces of portfolio management. Discretization strategies may lead to oversimplification and scalability issues. Continuous RL frameworks, like actor-critic Deterministic Policy Gradient Algorithms \cite{silver2014deterministic, lillicrap2016continuous}, offer a solution, despite their training complexities.

This study aims to harmonize advanced RL solutions with the time-tested principles of Markowitz Portfolio Theory, forging a path towards a more integrated and effective portfolio construction methodology. The integration of deep learning with Markowitz Portfolio Theory aims to overcome the limitations of applying deep learning in isolation for portfolio construction. This innovative approach employs knowledge distillation to merge the computational prowess of deep learning with the strategic foundations of efficient portfolio management, enhancing portfolio construction performance.

To actualize this harmonization, our study adopts a two-phased approach, blending theoretical understanding with practical application, to forge a new frontier in sophisticated portfolio construction. Initially, portfolios are constructed using Markowitz theory, creating a foundation for the deep learning agent's supervised learning. Following this, the agent engages in active reinforcement learning within the investment environment. This approach, inspired by the success of knowledge distillation in training large language models \cite{Liu_2023}, aims to meld deep learning with established financial theories, fostering a sophisticated model for portfolio construction.

The structure of this article is as follows: Section 2 introduces the relevant work and findings related to this paper, along with the shortcomings of these results. Section 3 provides the background of this study, presenting the techniques and evaluation metrics employed, both from financial and technological perspectives. Section 4 details the methods used in this study, including supervised learning and deep learning approaches. Section 5 elaborates on the experimental setup, while Section 6 discusses the experimental results and analysis. Finally, Section 7 concludes the paper with a summary and conclusions.

\section{Background}

\subsection{Markowitz model}
The seminal work of \cite{markowitz1968portfolio} laid the foundation for modern portfolio theory with the introduction of the classic Markowitz portfolio theory. This theory is predicated on the assumption that investors are risk-averse, seeking to minimize uncertainty and potential losses while maximizing expected returns for a given level of risk.

Central to the Markowitz model is the efficient frontier, which represents a set of optimal portfolios that either maximize expected return for a specified level of risk or minimize risk for a given expected return. Portfolios on the efficient frontier are considered optimally balanced.

The Markowitz portfolio model is mathematically formulated as an optimization problem:
\begin{align*}
\text{Maximize:} \quad & E(R_p) = \sum_{i=1}^{n} w_i E(R_i) \\
\text{Subject to:} \quad & \sum_{i=1}^{n} w_i = 1, \quad w_i \geq 0 \\
& \text{Var}(R_p) = \sum_{i=1}^{n} \sum_{j=1}^{n} w_i w_j \sigma_{ij},
\end{align*}
where $E(R_p)$ denotes the expected return of the portfolio, $w_i$ represents the weight of asset $i$ in the portfolio, $E(R_i)$ is the expected return of asset $i$, $\text{Var}(R_p)$ is the variance of the portfolio's return, indicative of risk, $\sigma_{ij}$ is the covariance between the returns of assets $i$ and $j$. The constraint $\sum_{i=1}^{n} w_i = 1$ ensures the portfolio is fully invested. The non-negativity condition $w_i \geq 0$ prohibits short selling. $n$ is the number of assets in the portfolio.

This optimization framework facilitates the identification of the efficient frontier. Portfolios below this frontier are sub-optimal as they fail to maximize expected return for their associated risk level, thereby not fully exploiting the potential of the invested capital.


The efficient frontier remains a pivotal concept in Markowitz's theory, depicted as a curve on a risk-return graph with risk (standard deviation of returns) on the x-axis and expected return on the y-axis. Portfolios on this curve are deemed efficient, as they offer an optimal trade-off between risk and return. Conversely, portfolios below the efficient frontier are sub-optimal, providing lower returns for equivalent levels of risk.

\subsection{Markov Decision Processes and DDPG}

Markov Decision Processes (MDPs) provide a mathematical framework for decision-making in stochastic control processes, with applications ranging from robotics to economics. MDPs, which originated in the 1950s and were developed by Howard and Puterman, incorporate actions and rewards into Markov chains, enabling decision-making and incentivization. In an MDP, a system in state \(s\) transitions to state \(s'\) with reward \(R_a(s, s')\) upon action \(a\), following the Markov property that future states depend only on the current state and action.

We frame a trading task as an MDP defined by \((S, A, P, R, \gamma)\), where \(S\) and \(A\) represent the state and action spaces, \(P(s'|s, a)\) is the transition probability, \(R(s, a, s')\) is the reward function, and \(\gamma\) is the discount factor. The agent's goal is to find a policy \(\pi(s_t|a_t)\) that maximizes the discounted cumulative return \( R = \sum_{t=0}^{T} \gamma^t R(s_t, a_t, s_{t+1}) \), transforming historical data into a market environment for strategy optimization.

The state \(s\) includes market information such as balance, shares, OHLCV data, technical indicators, and sentiment data, while the action \(a\) encompasses permissible trading actions. The reward \(R(s, a, s')\) is typically based on portfolio value change, log return, or the Sharpe ratio.

In cases of limited observability, we employ Partially Observable MDPs (POMDPs) with Hidden Markov Models (HMMs) to account for unobservable state sequences, using techniques like the off-policy Recurrent Deterministic Policy Gradient (RDPG) algorithm and LSTM networks to process partial observations.

However, the simulation-to-reality gap must be acknowledged, as training and testing on historical data may not accurately reflect live market conditions. This gap is exacerbated in financial RL by the low signal-to-noise ratio, survivorship bias, and the risk of overfitting during backtesting.

The Deep Deterministic Policy Gradient (DDPG) algorithm, introduced by Lillicrap et al., is a milestone in RL for continuous action spaces. As an actor-critic, model-free algorithm, DDPG operates off-policy with deep function approximators, adeptly handling high-dimensional spaces.

DDPG employs two neural networks: the actor network $\pi(s|\theta^\pi)$, mapping states to actions, and the critic network $Q(s, a|\theta^Q)$, evaluating actions by computing the Q-value function. The actor proposes the optimal action for a state, while the critic assesses the action's quality.

Experience replay is a key DDPG feature, storing agent experiences $(s_t, a_t, r_t, s_{t+1})$ in a replay buffer $R$ to sample random mini-batches for training, enhancing learning stability by decorrelating consecutive steps.

There are two arget networks: the actor and critic networks are exploited in the DDPG algorithm.

DDPG's suitability for complex, continuous action spaces, like financial portfolio optimization, is enhanced by knowledge distillation, a technique for training compact "student" models from sophisticated "teacher" models, introduced by Hinton et al. This improves learning and generalization while maintaining computational efficiency, vital for deploying models in resource-constrained environments.

\subsection{Knowledge Distillation}
Knowledge distillation involves a teacher model with a temperature-modified softmax output layer $
q_i = \frac{\exp(z_i/T)}{\sum_j \exp(z_j/T)}
$
where $z_i$ are logits, $T$ is the temperature, and $q_i$ are the softened probabilities. A higher $T$ yields a softer probability distribution.

The student model learns to match both true labels and softened probabilities from the teacher. The objective function combines cross-entropy loss with respect to true labels and an additional term for the softened probabilities:

\[
L = H(y, \sigma(z_{\text{student}})) + \lambda T^2 H(\sigma(z_{\text{teacher}}/T), \sigma(z_{\text{student}}/T))
\]

where $H$ is cross-entropy, $y$ are true labels, $\sigma$ is the softmax function, and $\lambda$ balances the two terms.

Knowledge distillation's efficiency is practical for scenarios requiring lightweight models, such as mobile or embedded systems. It enables complex model emulation with remarkable efficiency, broadening access to advanced machine learning for various applications, including financial investments.

\subsection{Evaluation Metrics}
In the realm of portfolio management, a variety of metrics are employed to assess performance and risk. These metrics are essential for comprehensive portfolio analysis and informed decision-making. Below is a summary of key performance indicators:

\textbf{Total Return} ($R_{\text{total}}$) quantifies the aggregate percentage return over the investment period. It is defined as \cite{herold2007total}:
$$
R_{\text{total}} = \left( \frac{P_{\text{final}} - P_{\text{initial}}}{P_{\text{initial}}} \right) \times 100
$$
where $P_{\text{final}}$ and $P_{\text{initial}}$ represent the final and initial portfolio values, respectively.

\textbf{Annualized Return} ($R_{\text{annualized}}$) provides the annual equivalent return, calculated as \cite{fama1990stock}:
$$
R_{\text{annualized}} = \left(1 + R_{\text{total}}\right)^{\frac{1}{n}} - 1
$$
with $n$ denoting the number of years.

\textbf{Sharpe Ratio} ($S$), introduced by William Forsyth Sharpe, measures risk-adjusted return \cite{sharpe1994sharpe}:
$$
S = \frac{R_{\text{portfolio}} - R_{\text{risk-free}}}{\sigma_{\text{portfolio}}}
$$
Here, $R_{\text{portfolio}}$ is the portfolio return, $R_{\text{risk-free}}$ the risk-free rate, and $\sigma_{\text{portfolio}}$ the standard deviation of the portfolio's excess return.

\textbf{Maximum Drawdown} ($D_{\text{max}}$) captures the largest peak-to-trough decline in portfolio value \cite{magdon2004maximum}:
$$
D_{\text{max}} = \max_{\tau}( \max_{t \leq \tau}(V_t) - V_{\tau} )
$$
where $V_t$ is the portfolio value at time $t$.

\textbf{Sortino Ratio} differentiates itself from the Sharpe Ratio by focusing solely on downside volatility \cite{rollinger2013sortino}:
$$
\text{Sortino Ratio} = \frac{R_p - R_f}{\sigma_d}
$$
with $R_p$ as the portfolio return, $R_f$ the risk-free rate, and $\sigma_d$ the standard deviation of negative asset returns.

\textbf{Beta} ($\beta$) gauges the volatility of a portfolio relative to the overall market \cite{fabozzi1977stability}:
$$
\beta = \frac{\text{Cov}(R_p, R_m)}{\text{Var}(R_m)}
$$
where $R_p$ and $R_m$ are the portfolio and market returns, respectively.

\textbf{Alpha} ($\alpha$) measures the active return on an investment compared to a market index \cite{fabozzi1977stability}:
$$
\alpha = R_p - (R_f + \beta(R_m - R_f))
$$

\textbf{Information Ratio} quantifies the excess returns per unit of risk against a benchmark \cite{goodwin1998information}:
$$
\text{Information Ratio} = \frac{R_p - R_b}{\text{Tracking Error}}
$$
where $R_b$ is the benchmark return, and the tracking error is the standard deviation of the portfolio-benchmark return differential.

\textbf{Calmar Ratio} compares the annualized return to the maximum drawdown \cite{sencomparative}:
$$
\text{Calmar Ratio} = \frac{\text{Annualized Return}}{D_{\text{max}}}
$$

\textbf{Win Rate} is the proportion of profitable trades to the total number of trades \cite{griffin1984different}:
$$
\text{Win Rate} = \frac{\text{Number of Profitable Trades}}{\text{Total Number of Trades}}
$$

\textbf{Profit/Loss Ratio} represents the average profit of winning trades against the average loss from losing trades \cite{lee2007economic}:
$$
\text{Profit/Loss Ratio} = \frac{\text{Average Profit of Winning Trades}}{\text{Average Loss of Losing Trades}}
$$

\textbf{Volatility} ($\sigma$) reflects investment risk through the standard deviation of portfolio returns \cite{shiller1992market}:
$$
\sigma = \sqrt{\text{Var}(R_p)}
$$

Each metric offers a distinct lens through which to view the investment portfolio's performance and risk profile, thereby facilitating a multifaceted evaluation of its efficacy.

\section{Proposed Method}
\begin{algorithm*}
\small
\renewcommand{\baselinestretch}{0.8}
\caption{Knowledge Distilled DDPG Algorithm}
\begin{algorithmic}[1]
\State Initialize actor network $\pi(s | \theta^{\pi})$ and critic network $Q(s, a | \theta^{Q})$
\State Initialize target networks $\pi'$ and $Q'$ with weights $\theta^{\pi'} \gets \theta^{\pi}, \theta^{Q'} \gets \theta^{Q}$
\State Initialize replay buffer $\mathcal{R}$
\State Pretrain $\pi(s | \theta^{\pi})$ using knowledge distillation:
\newcommand{\Indent}{\hspace*{0.5cm}} 
    \State Extract policy knowledge from a pre-trained Markowitz model
    \State Adapt the extracted knowledge to fit the DDPG actor network
    \State Train the actor network $\pi(s | \theta^{\pi})$ using this adapted knowledge
\newcommand{\EndIndent}{\par}
\For{episode $= 1, M$}
    \State Initialize a random process $\mathcal{N}$ for action exploration
    \State Receive initial observation state $s_1$
    \For{t $= 1, T$}
        \State Select action $a_t = \pi(s_t | \theta^{\pi}) + \mathcal{N}_t$ according to the current policy and exploration noise
        \State Execute action $a_t$ and observe reward $r_t$ and new state $s_{t+1}$
        \State Store transition $(s_t, a_t, r_t, s_{t+1})$ in $\mathcal{R}$
        \State Sample a random minibatch of $N$ transitions $(s_i, a_i, r_i, s_{i+1})$ from $\mathcal{R}$
        \State Set $y_i = r_i + \gamma Q'(s_{i+1}, \pi'(s_{i+1} | \theta^{\pi'}) | \theta^{Q'})$
        \State Update critic by minimizing the loss: $L = \frac{1}{N} \sum_i(y_i - Q(s_i, a_i | \theta^{Q}))^2$
        \State Update the actor policy using the sampled policy gradient:
        \[
        \nabla_{\theta^{\pi}} J \approx \frac{1}{N} \sum_i 
        \nabla_a Q(s, a | \theta^{Q})|_{s=s_i, a=\pi(s_i)} 
        \nabla_{\theta^{\pi}} \pi(s | \theta^{\pi})|_{s_i}
        \]
        \State Update the target networks:
        \[
        \theta^{\pi'} \gets \tau \theta^{\pi} + (1 - \tau) \theta^{\pi'}, \quad \theta^{Q'} \gets \tau \theta^{Q} + (1 - \tau) \theta^{Q'}
        \]
    \EndFor
\EndFor
\end{algorithmic}
\end{algorithm*}

Integrating knowledge distillation into the initial phase of Reinforcement Learning (RL) with the Deep Deterministic Policy Gradient (DDPG) approach represents a sophisticated fusion of machine learning techniques for financial portfolio management. This method entails the transfer of intricate decision-making expertise from a well-trained 'teacher' model to a nascent 'student' model, namely the DDPG. The objective is to endow the DDPG with advanced insights into investment strategies before it undergoes specialized training on financial datasets.

The procedure begins with the development of an advanced teacher model, skilled in portfolio optimization and knowledgeable about various financial scenarios, to generate a robust set of investment strategies. The teacher model's insights are then conveyed to the DDPG student model, aligning the student's policy function, $\pi_{\theta}(a|s)$, with that of the teacher, where $\theta$ represents the parameters of the student model, $a$ denotes the action, and $s$ the state.

The knowledge transfer is achieved by training the student model to replicate the teacher's outputs, guiding the DDPG to emulate decisions such as portfolio allocations in particular market conditions. During this training phase, the DDPG's parameters are fine-tuned to closely resemble the teacher model's decisions, with the goal of minimizing the divergence between their outputs using a loss function, $L(\theta)$, typically employing measures like the Kullback-Leibler divergence or Mean Squared Error.

By undergoing initial training via knowledge distillation, the DDPG acquires a foundational layer of financial acumen from the teacher model's extensive experience. This enhances the efficiency of its subsequent training on specific financial data and accelerates its advancement toward optimal portfolio strategies. The incorporation of knowledge distillation in initializing DDPG algorithms signifies an innovative convergence of machine learning paradigms, setting the stage for more proficient and effective solutions in the realm of financial portfolio management.

\subsection{Supervised Learning Stage}
In this segment of our research, we aim to establish foundational groundwork for the Deep Deterministic Policy Gradient (DDPG) agent through a supervised pre-training regimen, anchored in the esteemed principles of Markowitz's portfolio theory. Our study's overarching objective is to construct an investment portfolio optimized for maximum return yields while concurrently minimizing associated risks.

During the Supervised Learning Stage, our primary goal is to train the DDPG actor network to replicate the decision-making process of the Markowitz model, thereby effectively learning optimal portfolio strategies under various market conditions. The Mean Squared Error (MSE) loss function is critical in this stage as it quantifies the discrepancy between the decisions made by the DDPG actor network and the optimal decisions suggested by the Markowitz model.

   $$L(\theta) = \frac{1}{N} \sum_{i=1}^{N} (y_i - \hat{y}_i(\theta))^2$$

where:
\begin{itemize}
    \item $L(\theta)$ is the loss function dependent on the parameters $\theta$ of the DDPG actor network.
    \item $N$ is the number of samples in the minibatch.
    \item $y_i$ is the target output from the Markowitz model for the $i$-th sample, representing the optimal action.
    \item $\hat{y}_i(\theta)$ is the output of the DDPG actor network for the $i$-th sample, representing the action chosen by the network.
\end{itemize}

Initially, we generate a dataset based on the Markowitz model, simulating optimal investment portfolios under various market conditions. This dataset encompasses asset allocation, projected returns, and risk assessments. Neural network architectures are then employed to preprocess this dataset, ensuring compatibility with the input requirements of the DDPG actor network and providing exposure to a diverse array of market scenarios.

Subsequently, the refined dataset serves as a platform for the supervised learning phase of the DDPG actor network. This phase is instrumental in allowing the network to effectively emulate the decision-making processes inherent in the Markowitz model, thereby internalizing its strategic methodologies for portfolio management. This pre-training phase is pivotal as it not only instills a baseline strategy based on well-established financial theories but also enhances the DDPG agent with robust financial acumen, essential for navigating real-market environments and making prudent decisions.

Moreover, this approach significantly accelerates the DDPG agent's learning curve by endowing it with pre-learned patterns and strategies, greatly reducing the need for extensive data and time expenditure during subsequent reinforcement learning stages. Throughout this pre-training phase, our methodology integrates traditional portfolio management theories with advanced machine learning paradigms. This integration not only enriches the agent's repository of decision-making strategies but also maintains a harmonious balance between exploiting established strategies and exploring new market opportunities. Through this intricate blend, our model aims to advance the frontiers of portfolio management, merging traditional financial wisdom with innovative algorithmic finesse.

\subsection{Reinforcement Learning Stage}
Following the pretraining phase, the Deep Deterministic Policy Gradient (DDPG) agent enters the reinforcement learning (RL) phase, aiming to optimize financial portfolio performance. In this phase, the RL agent interacts with an environment characterized by fluctuating market conditions, including stock prices, trading volumes, and various financial indicators. The agent is endowed with a range of actions such as buying, selling, or holding different asset classes, with these decisions being heavily influenced by current market dynamics.

To assess the agent's performance, we have devised a comprehensive reward function that incorporates key financial metrics, including return on investment (ROI) and risk-adjusted returns. This function provides a complete measure of portfolio performance. The agent's training involves a cyclical process of observing market states, executing policy-driven decisions, receiving feedback through rewards, and iteratively refining its policy to enhance decision-making capabilities.

The DDPG algorithm's implementation has been carefully adapted to fine-tune the agent's policy, balancing the exploration of new strategic options with the exploitation of known profitable approaches. This policy improvement aligns with the standard DDPG training protocol, with specific adjustments made to accommodate the unique aspects of the financial market environment.

Backtesting with historical market data serves as a stringent method to assess the agent's performance. This critical phase evaluates the effectiveness of the learned strategies and their applicability in real-market scenarios. The goal of this thorough evaluation is to confirm the model's practicality and resilience.

Ultimately, the RL phase is crucial for refining and enhancing the investment strategies developed during the pretraining phase. This phase leverages the foundational concepts of Markowitz portfolio theory while adapting to the complex and dynamic nature of actual financial markets. This two-pronged approach ensures that the DDPG agent is well-anchored in established financial theories and adept at adjusting to the intricacies of live market conditions.

\section{Numerical Experiments}
\subsection{Experiment Setup}
In the initial phase of our experimental methodology, we focused on feature engineering for the neural network's input. This involved using stock closing prices, financial instrument identifiers, and computed asset allocations, tailored to meet the requirements of the Deep Deterministic Policy Gradient (DDPG) neural network. Following feature engineering, we applied knowledge distillation during pre-training to enhance the network's ability to decipher stock market complexities and dynamics.

The DDPG agent underwent extensive training with the engineered data, aiming to develop a model with a deep understanding and effective performance in the stock market environment.

During the validation stage, we fine-tuned critical hyperparameters such as the learning rate and the number of training episodes using validation data to optimize the agent's performance metrics. In the trading phase, we evaluated the agent's performance against trading data, with ongoing training to maintain adaptability to the market's changing conditions.

Our experimental setup included thorough data preprocessing and feature selection, utilizing historical daily stock prices from the Dow Jones 30 between January 1, 2009, and September 30, 2018. We normalized stock prices to ensure data consistency and to facilitate efficient learning by the neural network. Our feature engineering aimed to capture the market's complex dynamics.

The pre-training of the DDPG network was characterized by the use of knowledge distillation techniques. We integrated the distilled knowledge from the Markowitz model, which calculated optimal asset allocation ratios, into the DDPG network. This integration aimed to enable the network to effectively mimic investment strategies based on Markowitz's theory.

Our experimental approach was structured into three main stages: training, validation, and trading. Each stage played a crucial role in the comprehensive development and empirical evaluation of our model. We used a range of financial metrics, including Total Return, Annualized Return, and risk-adjusted measures like the Sharpe Ratio, to assess the model's performance in portfolio management. These metrics provided valuable insights into the model's effectiveness.

\subsection{Results} 

\begin{table*}[h]
\centering
\caption{Consolidated Performance Comparison of Investment Strategies. The table presents a comprehensive comparison of various investment strategies across multiple performance metrics. Abbreviations: DJI - Dow Jones Industrial, BAH - Buy And Hold, BCRP - Best Constant Rebalanced Portfolio, CRP - Constant Rebalanced Portfolio, MVO - Mean-Variance Optimization, EG - Exponential Gradient, UP - Universal Portfolio, ONS - Online Newton Step, SP - Stochastic Portfolio, ACO - Ant Colony Optimization, PSO - Particle Swarm Optimization, CWMR - Confidence Weighted Mean Reversion, OLMAR - On-Line Moving Average Reversion, Bk - Benchmark Strategy k, BNN - Benchmark Strategy Neural Network, CORN - Correlation Driven Nonparametric Learning, DDPG - Deep Deterministic Policy Gradient, MKD - Markowitz Knowledge Distillation, KDD - Knowledge Distillation DDPG. Performance metrics: TR - Total Return, AR - Annualized Return, Sharpe - Sharpe Ratio, MD - Max Drawdown, SR - Sortino Ratio, IR - Information Ratio, CR - Calmar Ratio, WR - Win Rate, PLR - Profit/Loss Ratio.}
\label{tab:consolidated_investment_strategies}
\footnotesize	 
\setlength{\tabcolsep}{2pt} 
\renewcommand{\arraystretch}{0.8} 
\setlength{\tabcolsep}{3.2pt} 
\begin{tabular}{c|c|c c c c c c c c c c c c}
\hline
\textbf{Categories} & \textbf{Strategy} & \textbf{TR} & \textbf{AR} & \textbf{Sharpe} & \textbf{MD} & \textbf{SR} & \textbf{Beta} & \textbf{Alpha} & \textbf{IR} & \textbf{CR} & \textbf{WR} & \textbf{PLR} & \textbf{Volatility} \\
\hline
\multirow{6}{*}{Baselines} & DJI & 51.41\% & 16.91\% & 1.21 & -11.58\% & 0.12 & 1.00 & -0.03 & - & 1.46 & 55.59\% & 1.03 & 0.75\% \\
& BAH & 53.11\% & 17.40\% & 1.25 & -10.70\% & 0.13 & 0.98 & 0.85 & 0.01 & 1.63 & 56.33\% & 1.01 & 0.74\% \\
& BCRP & 63.31\% & 20.31\% & 1.41 & -10.43\% & 0.14 & 0.98 & 3.75 & 0.05 & 1.95 & 57.46\% & 0.99 & 0.77\% \\
& CRP & 53.07\% & 17.39\% & 1.27 & -10.76\% & 0.13 & 0.96 & 1.16 & 0.01 & 1.62 & 56.48\% & 1.01 & 0.73\% \\
& Markowitz & 69.43\% & 21.96\% & 1.27 & -14.10\% & 0.13 & 0.99 & 5.12 & 0.03 & 1.56 & 54.55\% & 1.09 & 0.93\% \\
& MVO & 61.92\% & 19.92\% & 1.35 & -11.58\% & 0.14 & 0.91 & 4.51 & 0.02 & 1.72 & 55.52\% & 1.06 & 0.79\% \\
\hline

\multirow{8}{*}{Follow the winner} & EG & 53.13\% & 17.42\% & 1.27 & -10.76\% & 0.13 & 0.96 & 1.20 & 0.01 & 1.62 & 56.42\% & 1.01 & 0.73\% \\
& UP & 47.09\% & 15.66\% & 1.13 & -10.60\% & 0.11 & 0.95 & -0.41 & -0.02 & 1.48 & 56.87\% & 0.97 & 0.74\% \\
& ONS & 53.22\% & 17.45\% & 1.27 & -10.72\% & 0.13 & 0.96 & 1.18 & 0.01 & 1.63 & 56.27\% & 1.02 & 0.73\% \\
& SP & 26.76\% & 9.35\% & 0.73 & -15.08\% & 0.08 & 0.72 & -2.85 & -0.07 & 0.62 & 55.97\% & 0.93 & 0.65\% \\
& ACO & 98.16\% & 29.41\% & 1.05 & -29.85\% & 0.10 & 1.05 & 11.60 & 0.04 & 0.99 & 52.69\% & 1.10 & 1.63\% \\
& Annealing & 55.59\% & 18.13\% & 1.25 & -11.62\% & 0.13 & 0.99 & 1.37 & 0.02 & 1.56 & 56.72\% & 0.99 & 0.78\% \\
& PSO & 72.60\% & 22.84\% & 1.56 & -11.86\% & 0.15 & 0.97 & 6.41 & 0.07 & 1.93 & 57.76\% & 1.01 & 0.78\% \\
& Genetic & 40.72\% & 13.74\% & 1.03 & -12.56\% & 0.11 & 0.84 & -0.50 & -0.03 & 1.09 & 56.87\% & 0.95 & 0.70\% \\
\hline

\multirow{4}{*}{Follow the loser} & Anticor & 56.93\% & 21.62\% & 1.56 & -8.27\% & 0.16 & 0.90 & 5.36 & 0.03 & 2.61 & 56.63\% & 1.05 & 0.74\% \\
& PAMR & 86.50\% & 26.48\% & 1.84 & -10.14\% & 0.18 & 1.00 & 9.61 & 0.22 & 2.61 & 58.81\% & 1.02 & 0.76\% \\
& CWMR & 69.22\% & 21.93\% & 1.57 & -10.32\% & 0.16 & 0.98 & 5.31 & 0.13 & 2.12 & 58.21\% & 1.00 & 0.74\% \\
& OLMAR & 96.47\% & 29.19\% & 1.24 & -13.08\% & 0.13 & 1.32 & 3.99 & 0.04 & 2.23 & 51.35\% & 1.19 & 1.31\% \\
\hline

\multirow{3}{*}{Pattern Matching} & Bk & 43.67\% & 16.17\% & 0.98 & -15.40\% & 0.10 & 1.08 & -2.50 & -0.00 & 1.05 & 54.99\% & 1.00 & 0.90\% \\
& BNN & 45.23\% & 16.69\% & 1.14 & -11.17\% & 0.12 & 1.04 & -1.29 & -0.00 & 1.49 & 55.81\% & 1.01 & 0.79\% \\
& CORN & 57.93\% & 19.75\% & 1.49 & -11.46\% & 0.15 & 1.00 & 0.03 & 0.00 & 1.72 & 56.94\% & 1.03 & 0.70\% \\
\hline

\multirow{2}{*}{Others} & M0 & 53.03\% & 17.39\% & 1.27 & -10.76\% & 0.13 & 0.96 & 1.18 & 0.01 & 1.62 & 56.42\% & 1.01 & 0.73\% \\
&T0 & 33.58\% & 11.53\% & 0.54 & -16.66\% & 0.06 & 1.09 & -6.91 & -0.01 & 0.69 & 49.85\% & 1.12 & 1.29\% \\
\hline

\multirow{3}{*}{AI models} & DDPG & 45.08\% & 15.06\% & 0.98 & -16.41\% & 0.10 & 0.69 & 3.44 & -0.01 & 0.92 & 54.63\% & 1.02 & 0.83\% \\
&MKD & 58.69\% & 18.99\% & 1.31 & -10.28\% & 0.13 & 0.95 & 2.87 & 0.02 & 1.85 & 56.04\% & 1.04 & 0.77\% \\
&KDD & \textbf{138.38\%} & \textbf{38.74\%} & \textbf{2.03} & -11.46\% & \textbf{0.21} & 1.03 & \textbf{21.31} & \textbf{0.11} & \textbf{3.38} & 55.37\% & 1.18 & 0.99\% \\
\hline
\end{tabular}
\smallskip 
\raggedright 

\end{table*}

In the domain of portfolio management, a plethora of investment strategies have been developed, each with distinct perspectives on risk, return, and market dynamics. Our paper conducts a comparative analysis of these strategies, focusing on performance metrics such as win rate, profit/loss (P/L) ratio, and volatility. We examine traditional models like the Dow Jones Industrial (DJI) and Markowitz Portfolio Theory, as well as advanced AI-driven methods, including the Deep Deterministic Policy Gradient (DDPG) and its enhanced variant, the KDD (Knowledge Distilled DDPG) model.

The DJI serves as a benchmark index, indicative of market trends and a standard for performance comparison. The Markowitz model, a cornerstone of modern portfolio theory, advocates for diversification to balance risk and return. The DDPG, leveraging reinforcement learning (RL), dynamically adjusts portfolio allocations in response to market conditions.

Strategies such as the Ant Colony Optimization (ACO) Optimized Portfolio and Simulated Annealing Optimized Portfolio employ computational algorithms inspired by natural phenomena to identify optimal asset allocations. The Anticor Model and the Buy and Hold (BAH) strategy offer contrasting approaches to market timing and long-term investment, respectively.

Other strategies, including the Constant Rebalanced Portfolio (CRP), the Capital Growth Model (CWMR), and Online Portfolio Selection (OLPS) models like OLMAR and ONS, present diverse methods for asset reallocation and online decision-making in light of market fluctuations. The integration of machine learning and optimization models, such as the Best k Model, Best NN Model, and Genetic Optimized Portfolio, underscores the increasing role of AI and computational intelligence in financial decision-making.

Each strategy encapsulates a unique investment philosophy and methodology, providing a spectrum of approaches for navigating financial markets. Our analysis offers a holistic view of the portfolio management landscape, delineating the advantages, limitations, and applicability of each strategy under varying market conditions.

The KDD model demonstrates exceptional performance, with a Total Return of \textbf{138.38\%}, outshining all other strategies, including the DJI and Markowitz portfolio, which yielded returns of 51.41\% and 69.43\%, respectively.

Moreover, the model achieves an Annualized Return of \textbf{38.74\%}, surpassing all competitors. This metric, which normalizes returns over time, facilitates a more equitable comparison across different periods. The model's lead in this area underscores its consistent high-return generation capability.

In terms of risk-adjusted returns, the model attains the highest Sharpe Ratio of \textbf{2.03}, indicating not only higher returns but also a superior risk-return balance. A Sharpe Ratio above 2 is exceptional, suggesting that the model's performance is not merely a function of increased risk but effective strategy and decision-making.

The KDD model's superior performance across Total Return, Annualized Return, and Sharpe Ratio confirms its robustness as an investment strategy. Its ability to deliver significantly higher returns while maintaining an excellent risk-return profile positions it as a noteworthy advancement in algorithm-driven portfolio management.

The model's Max Drawdown of -11.46\% is moderate and comparable to strategies like the DJI and MVO. Max Drawdown assesses the largest drop from peak to trough in a portfolio's value before a new peak is achieved. Although not the lowest, the model's drawdown reflects a managed risk level, particularly given its high return profile, indicating a strategy that pursues high returns without excessive risk exposure.

The model also achieves the highest Sortino Ratio of \textbf{0.21}, signifying efficient generation of returns above the risk-free rate while minimizing downside risks. The Sortino Ratio, which concentrates on negative volatility, suggests that the strategy effectively captures upward market trends while protecting against significant downturns.

With a Beta of 1.03, the model exhibits a market-level volatility, mirroring market movements. This Beta, combined with other metrics, implies that the model is well-tuned to capitalize on market trends without deviating significantly from market risk levels.

The KDD model represents a balanced approach in portfolio management, achieving high returns without undue downside risk exposure. Its high Sortino Ratio, moderate Max Drawdown, and market-aligned Beta underscore its potential as a reliable and effective tool for algorithm-driven investment strategies, offering a compelling option for investors seeking robust returns with a controlled risk profile.

The model's Alpha, the highest among all compared strategies, underscores its market outperformance. An Alpha of 21.31 indicates that the KDD model employs a potent strategy capable of generating superior returns, even after adjusting for market volatility (Beta). This high Alpha reflects the model's adept use of predictive insights and market inefficiencies, leveraging the advanced capabilities of DDPG with knowledge distillation.

The Information Ratio of the model, while not the highest in the dataset, is commendable. This metric assesses the consistency and predictability of excess returns relative to a benchmark, considering the level of risk incurred. The positive Information Ratio observed with the KDD model suggests that the strategy consistently outperforms the market with a reasonable increment of risk, indicating the model's capability to generate returns that are not only high but also reliable and repeatable over time.

Furthermore, the model's Calmar Ratio stands as the highest among the strategies analyzed. The Calmar Ratio, a performance metric, evaluates the risk-adjusted return of an investment strategy by focusing on the relationship between annualized return and maximum drawdown. The high Calmar Ratio associated with the KDD model indicates that the strategy provides substantial rewards for the risks undertaken, especially when considering potential temporary declines in value.

The KDD model exhibits a superior ability to generate high returns above the market average, maintain consistent performance over the benchmark, and offer an excellent balance between return and risk. This is evidenced by its leading position in Alpha and Calmar Ratio, highlighting the effectiveness of integrating knowledge distillation into the DDPG framework to enhance predictive power and decision-making efficiency in portfolio management. The model emerges as a potent and efficient tool within AI-driven investment strategies.

Although the win rate of the KDD model is robust, it is not the highest among the strategies. The win rate reflects the proportion of trades or investment decisions that yield a positive return. A win rate above 50\% indicates that the majority of the model's trading decisions are profitable, showcasing its effective decision-making capabilities. This win rate, while notable, also acknowledges the inherent uncertainties and complexities of financial markets that sophisticated models like the KDD must navigate.

The model's profit/loss ratio is nearly the highest in the dataset, marginally trailing the OLMAR Model. This ratio compares the average profit of winning trades to the average loss of losing trades. A ratio greater than 1, as demonstrated by the KDD model, signifies that the model's winning trades are on average more profitable than the losses from losing trades. This emphasizes the model's efficiency in not only securing wins but also in effectively managing and mitigating losses, which is vital for sustainable trading strategies.

The model presents moderate volatility, which is considered reasonable given the high returns it achieves. Volatility, a crucial measure of risk, indicates the variability of investment returns over time. The volatility of the KDD model reflects its dynamic response to market fluctuations, balancing the pursuit of high returns with risk management. This level of volatility is deemed acceptable, particularly in light of the model's high returns and sophisticated risk management strategies.

In summary, the KDD model demonstrates a well-balanced performance across win rate, profitability, and risk management. Its ability to maintain a high profit/loss ratio and a reasonable win rate, along with manageable volatility, underscores its effectiveness as an advanced investment strategy. These characteristics highlight the model's sophisticated design, which marries knowledge distillation with the DDPG algorithm, rendering it a formidable tool in the domain of algorithmic trading and portfolio management.

\begin{figure*}[htbp]
\centering
\includegraphics[width=\linewidth]{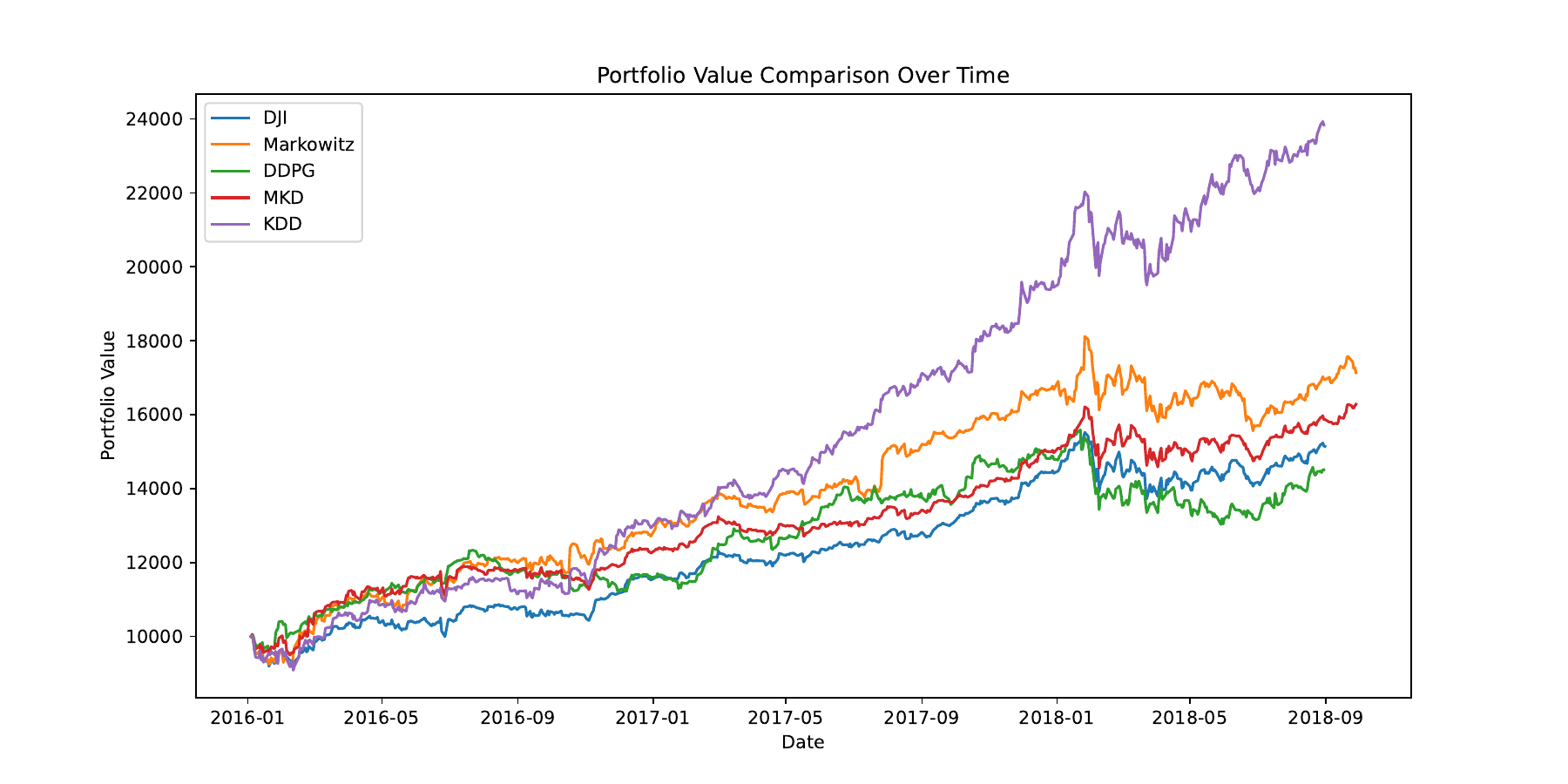}
\caption{Portfolio Value Comparison Over Time}
\label{fig:Portfolio Value Comparison Over Time}
\end{figure*}

In Figure \ref{fig:Portfolio Value Comparison Over Time}, the model demonstrates a consistent upward growth trajectory, which is a key indicator of a resilient and adaptive strategy in response to market dynamics. This steady increase in portfolio value highlights the model's ability to generate positive returns and its proficiency in managing the complexities inherent in financial markets.

Compared with other models such as "DJI," "Markowitz," "DDPG," and "Markowitz Knowledge Distilled," the "Knowledge Distilled DDPG" model exhibits competitive performance. Although it may not always lead in terms of absolute value gains, its consistency is a notable advantage. This characteristic is especially valuable in long-term investment strategies where reliability and the capacity to compound gains over time are paramount.

The "Knowledge Distilled DDPG" model also distinguishes itself through its stability, showing a lower volatility level relative to its peers. This trait suggests the implementation of an advanced risk management strategy that seeks a balance between potential returns and market uncertainties. Such stability is attractive to investors who are concerned with risk-adjusted returns.

Furthermore, the model's robust performance during both peak and trough phases of the market cycle reinforces its dependability. Its resilience in downturns and ability to leverage upturns reflect a finely-tuned approach that is sensitive to market fluctuations while aiming for sustained growth.

As depicted in the dataset, the "Knowledge Distilled DDPG" model emerges as a comprehensive investment strategy. Its consistent growth, comparative robustness, and stability amidst market volatility position it as an attractive option for investors aiming for long-term capital growth. The model's combination of reliable performance and effective risk management makes it a suitable component for diverse investment portfolios, appealing to those who seek a harmony between steady growth and cautious risk exposure.

\begin{figure*}[h]
\centering
\includegraphics[width=\linewidth]{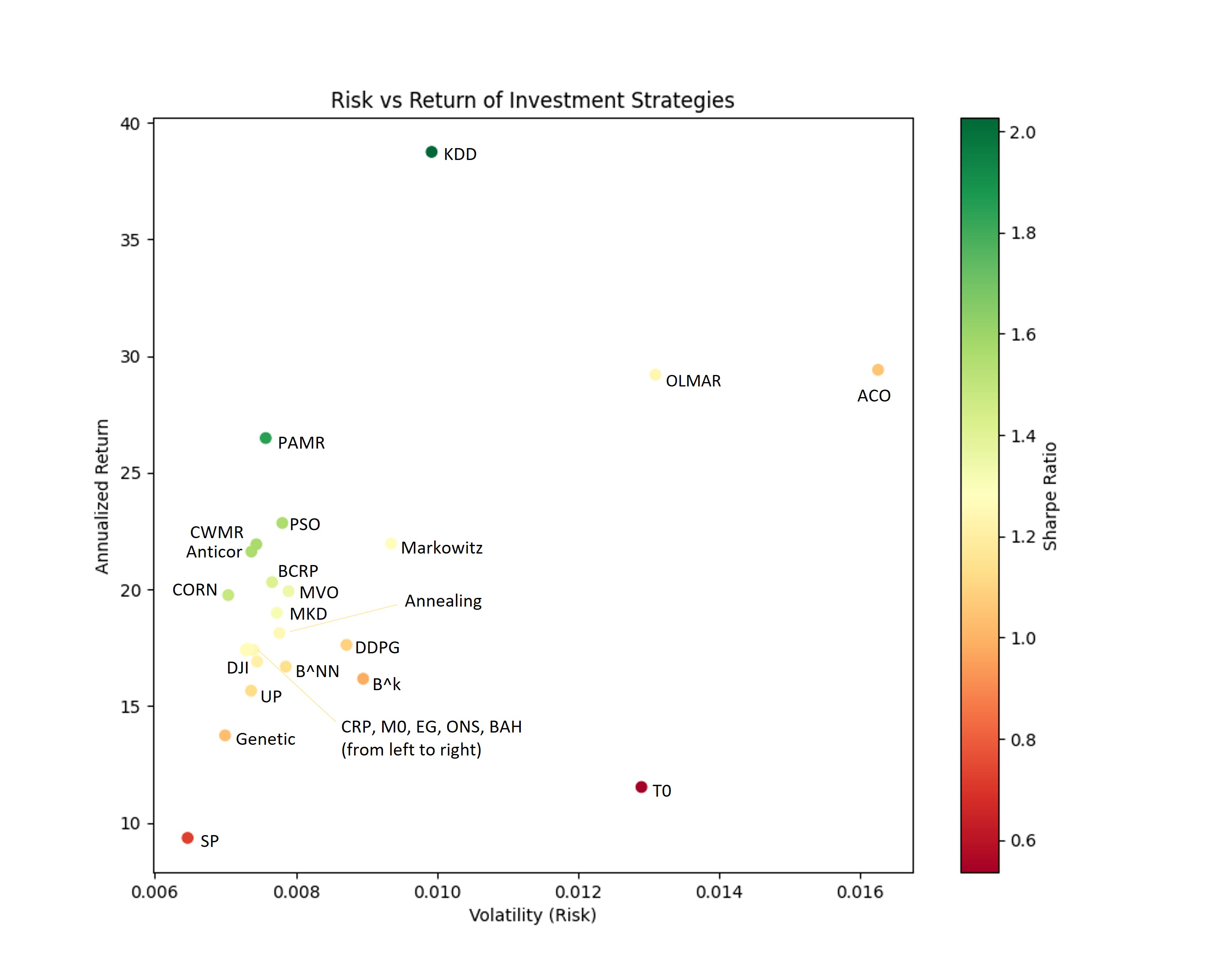}
\caption{\textbf{Risk vs Return of Investment Strategies}}
\label{fig:Risk vs Return of Investment Strategies}
\end{figure*}

In Figure \ref{fig:Risk vs Return of Investment Strategies}, the "Knowledge Distilled DDPG" model's performance within the realm of investment strategies underscores its remarkable efficiency and efficacy. The model stands out with an impressive annualized return of 38.74\%, significantly surpassing those of its counterparts. This high rate of return is indicative of a strategy that is not only adept at identifying lucrative market opportunities but also excels at capitalizing on them.

The "Knowledge Distilled DDPG" model exhibits notable volatility, recorded at 0.0099. Although this volatility is marginally higher compared to other models, it is essential to consider this within the context of risk-reward trade-offs. The elevated volatility may be indicative of the model's assertive strategy in maximizing returns, suggesting a calculated acceptance of increased risk that is designed to yield significantly higher returns.

The Sharpe Ratio, indicative of risk-adjusted performance, reinforces the model's efficacy. The model's exceptional returns, even after accounting for the heightened risk, suggest an efficient risk utilization to achieve substantial gains. This characteristic is particularly relevant for investors who are amenable to higher volatility in pursuit of greater potential rewards.

Within the investment strategy landscape, the "Knowledge Distilled DDPG" model signifies a shift towards a dynamic and assertive investment approach. Its performance highlights the potential of sophisticated investment strategies that embrace higher risks to deliver markedly higher returns. This model is especially attractive to investors who prioritize high returns and possess the capability to manage associated risk levels.

In essence, the "Knowledge Distilled DDPG" model strikes a refined balance between high risk and high reward, establishing a new standard in investment strategy performance. Its capacity to generate significantly higher returns, despite the increased volatility, renders it an appealing choice for investors targeting aggressive growth opportunities in their portfolios.

The model's performance is distinguished by its extraordinary total and annualized returns. With a total return of 138.38\%, it substantially surpasses its counterparts, signifying an efficacious capital appreciation strategy over the investment period. This figure encapsulates the cumulative return, encompassing capital gains and, if applicable, reinvested dividends.

The annualized return, at 38.74\%, is equally noteworthy. This metric, which adjusts the total return to an annual rate, offers a more transparent view of the model's performance over the investment duration. The "Knowledge Distilled DDPG" model's significantly high annualized return suggests a consistent outperformance of market averages, denoting a superior investment strategy, particularly when considering the compounding effect of returns over time.

The contrast between total and annualized returns is pivotal. While total return provides a snapshot of the investment's overall growth, annualized return sheds light on the strategy's consistency and dependability. The "Knowledge Distilled DDPG" model's impressive metrics in both areas emphasize its strength in achieving both short-term gains and long-term investment stability.

These outcomes are likely a result of the model's algorithmic foundation, which enables market condition adaptation, future trend forecasting, and investment allocation optimization. Such a strategy not only seizes market opportunities but also adeptly manages risks, as evidenced by the high return rates.

In conclusion, the "Knowledge Distilled DDPG" model demonstrates exceptional investment management capabilities, with its high total and annualized returns reflecting a profound grasp of market dynamics and effective strategy execution. Its analytical performance positions it as an exceedingly competent model for investors seeking significant portfolio growth.

\section{Conclusion}
In this study, we introduced the KDD (Knowledge Distillation DDPG) investment model, which has demonstrated significant insights into portfolio management. The model exhibited exceptional performance, achieving a total return of 138.38\% and an annualized return of 38.74\%, outperforming its counterparts. This remarkable performance is indicative of a sophisticated algorithmic framework that likely employs advanced machine learning techniques, predictive analytics, and effective use of real-time market data. The model's proficiency in discerning market dynamics suggests its capability to seize investment opportunities while adeptly managing risks.

The findings have profound implications for the domain of financial investment strategies. The high returns generated by the model position it as an invaluable asset for investors and financial institutions seeking robust, high-yield strategies. Its integration into investment decision-making could herald a shift towards more data-driven, informed financial planning and portfolio management, potentially transforming traditional practices.

Nonetheless, the study recognizes the model's limitations. Its performance under varying market conditions remains to be thoroughly evaluated, casting uncertainty on its adaptability and effectiveness in different economic climates. Additionally, the use of historical data to gauge performance may not be a reliable indicator of future market behavior, particularly in volatile or unpredictable financial scenarios.

Practically, the model harbors significant potential for application within the financial industry, including portfolio management, investment advising, and decision-making tools. Its analytical approach to investment could lead to improved returns and more sophisticated risk management strategies.

Future research should focus on assessing the model's performance across diverse economic cycles and geographic markets to gain a deeper understanding of its versatility and robustness. Enhancements could involve incorporating a wider array of data sources, such as global economic indicators and geopolitical events, to refine predictive accuracy. Moreover, the exploration of emerging technologies like artificial intelligence and blockchain may offer novel ways to bolster the security and efficiency of investment strategies.

In conclusion, the "Knowledge Distilled DDPG" model marks a significant step forward in optimizing investment strategies. Its promising returns underscore the importance of continued research and development to enhance its practicality and dependability in the ever-evolving and intricate financial market landscape.

\bibliographystyle{unsrt}  
\bibliography{references}

\end{document}